\documentclass[lettersize,journal]{IEEEtran}
\usepackage{amsmath,amsfonts,amssymb}
\usepackage{bm}
\usepackage{textcomp}
\usepackage{stfloats}
\usepackage{url}
\usepackage{verbatim}
\usepackage{graphicx}
\usepackage[shortlabels]{enumitem}
\usepackage{needspace}
\usepackage{balance}
\usepackage{comment}
\usepackage[table]{xcolor}
\usepackage{algorithm}
\usepackage{algpseudocode}
\usepackage{wasysym}
\usepackage{multirow}
\usepackage[numbers,sort&compress]{natbib}

\usepackage{blindtext}
\usepackage{wrapfig}
\usepackage{booktabs}
\usepackage{subfigure}
\usepackage{tcolorbox}
\tcbuselibrary{breakable}
\usepackage{soul}
\usepackage{colortbl}
\usepackage{listings}
\usepackage{tabularx}
\usepackage{adjustbox}
\usepackage{seqsplit}
\usepackage{xspace}
\usepackage{hyperref}
\usepackage[hyphenbreaks]{breakurl}

\definecolor{lightpink}{rgb}{1.0, 0.87, 0.87}
\definecolor{lightpurple}{rgb}{0.94, 0.85, 0.94}
\definecolor{lightgreen}{rgb}{0.88, 1.0, 0.88}
\definecolor{lightyellow}{rgb}{1.0, 1.0, 0.88}
\definecolor{lightblue}{rgb}{0.88, 0.94, 1.0}
\definecolor{lightgrey}{rgb}{0.8, 0.8, 0.8}
\definecolor{codegreen}{rgb}{0,0.6,0}
\definecolor{codegray}{rgb}{0.5,0.5,0.5}
\definecolor{codepurple}{rgb}{0.58,0,0.82}
\definecolor{backcolour}{rgb}{0.95,0.95,0.92}

\lstdefinestyle{mystyle}{
    backgroundcolor=\color{backcolour},
    commentstyle=\color{codegreen},
    keywordstyle=\color{magenta},
    numberstyle=\tiny\color{codegray},
    stringstyle=\color{codepurple},
    basicstyle=\footnotesize\ttfamily,
    breakatwhitespace=false,
    breaklines=true,
    captionpos=b,
    keepspaces=true,
    numbers=left,
    numbersep=5pt,
    showspaces=false,
    showstringspaces=false,
    showtabs=false,
    tabsize=3,
    frame=tb,
    aboveskip=2mm,
    belowskip=2mm
}
\begin{document}

\title{\textsc{GDBack}: A Group-Delay Backdoor Attack on Speaker Identification via Analytic All-Pass Filtering}

\author{Bicheng Xiong,
        Tao Li,
        Chenhui Cui,
        Zedong Xing,
        Qien Li,
        Weiping Wen,
        Rubing Huang,~\IEEEmembership{Senior Member,~IEEE},
        and Tsong Yueh Chen,~\IEEEmembership{IEEE Fellow,~IEEE}
\thanks{Bicheng Xiong and Zedong Xing are with Xiangtan University, Xiangtan 411105, China (e-mail: 202531630502@smail.xtu.edu.cn; 202431630378@smail.xtu.edu.cn).}%
\thanks{Chenhui Cui is with the University of Alberta, Edmonton, AB T6G 2R3, Canada (e-mail: chenhui.cui@ualberta.ca).}
\thanks{Qien Li and Weiping Wen are with Peking University, Beijing 100871, China (e-mail: liqien@pku.edu.cn; weipingwen@pku.edu.cn).}%
\thanks{Tao Li is with Beihang University, Beijing 100191, China (e-mail: abelli@buaa.edu.cn).}%
\thanks{Rubing Huang is with Macau University of Science and Technology, Macao SAR 999078, China (e-mail: rbhuang@must.edu.mo).}%
\thanks{
Tsong Yueh Chen is with Swinburne University of Technology, Melbourne 3122, Australia (e-mail: tychen@swin.edu.au).}%
\thanks{Tao Li and Weiping Wen are the corresponding authors.}
}

\maketitle

\begin{abstract}
Existing phase-based backdoor attacks on speaker identification establish cross-sample trigger consistency through shared absolute phase patterns. 
Meanwhile, other audio backdoor attacks have explored shared transformations that produce input-dependent triggers. 
However, whether phase-based trigger consistency can similarly arise from a shared transformation without imposing identical output phase patterns remains unclear.
We therefore explore a controllable phase transformation that can establish a consistent trigger cue while largely preserving the acoustic and linguistic characteristics of the original speech.
We propose \textsc{GDBack}, a poison-only group-delay backdoor attack that uses cascaded second-order analytic all-pass filters.
By applying the same transformation to different utterances, \textsc{GDBack} ideally preserves spectral magnitude while introducing a shared frequency-dependent phase response and group-delay contribution, allowing the output phases to remain dependent on the original speech.
Across LibriSpeech and VoxCeleb1 with ECAPA-TDNN, x-vector, and RawNet3, a single trigger configuration achieves an attack success rate (ASR) above 94\% in all six settings (mean 97.00\%), while maintaining benign accuracy close to that of clean models.
Controlled phase and magnitude interventions, inverse compensation, and parameter-sensitivity experiments link backdoor activation to the induced phase transformation and filter configuration.
Visual comparisons and speech quality and content assessments indicate that triggered speech largely preserves its structure, quality, and linguistic content.
Experiments involving signal and acoustic variations, including real over-the-air playback and recording, show that attack effectiveness varies across conditions and victim architectures.
These results show that cross-sample trigger consistency can be established at the transformation level without requiring poisoned utterances to share a common absolute phase pattern.
\end{abstract}


\section{Introduction}
\label{SEC:Introduction}

\IEEEPARstart{S}{PEAKER} recognition systems are widely used in biometric authentication and other identity-sensitive applications~\cite{ref1}.
Modern speaker recognition systems rely on deep speaker representations learned from speech data~\cite{ref2}.
However, practical training pipelines may incorporate third-party datasets, pretrained models, or external training resources, creating opportunities for training-time compromise~\cite{ref3,ref32,ref33}.
In a data-poisoning backdoor attack, an adversary manipulates a portion of the training data to associate a specific trigger with an attacker-selected target label, causing the trained model to behave normally on benign inputs but to produce the target prediction when the trigger is present~\cite{ref3}.
For speaker identification systems, such triggers can be embedded directly into speech recordings, making trigger construction an important factor in attack effectiveness and stealthiness.

Audio backdoor attacks have explored diverse trigger constructions, ranging from introducing external acoustic cues to transforming the original speech.
Koffas et al. employed inaudible ultrasonic signals as triggers for speech recognition systems~\cite{ref4}, while opportunistic attacks exploit ambient sounds that naturally occur in everyday environments~\cite{ref5}.
Other approaches modify properties of the speech itself, including padding operations in PaddingBack~\cite{ref6}, voice conversion in VCAttack~\cite{ref7}, emotional prosody in EmoBack~\cite{ref8}, and speaking-speed modifications in Speed Master~\cite{ref9}.
In particular, Speed Master and modulation-based methods such as ASMA and FSMA apply shared transformations to different utterances, generating input-dependent trigger cues without imposing identical output patterns~\cite{ref9,ref25}.

Phase information plays an important role in speech processing, including speech enhancement and recognition~\cite{ref34,ref43}.
Recent studies have explored its potential for constructing stealthy backdoor triggers in speaker recognition.
Specifically, PhaseBack injects predefined phase values into selected time-frequency components~\cite{ref10}, whereas StealthPhase replaces selected STFT phase regions with a phase pattern derived from reference speech~\cite{ref11}.
Despite their different implementations, both methods establish cross-sample trigger consistency by imposing shared absolute phase patterns on selected components of poisoned utterances.
Although these methods extend trigger construction to the phase domain, they have not investigated whether such consistency can be established through a shared phase transformation, as seen in other audio backdoor designs.
Our comparative experiments further show that their attack success rates vary substantially across datasets and speaker recognition architectures.
We therefore investigate a shared phase transformation as an alternative means of establishing cross-sample trigger consistency while retaining input-dependent output phases.
We also seek a controllable transformation that largely preserves the spectral magnitude, speech quality, and linguistic content of the original speech.

\begin{figure*}[t]
\centering
\includegraphics[width=0.96\textwidth]{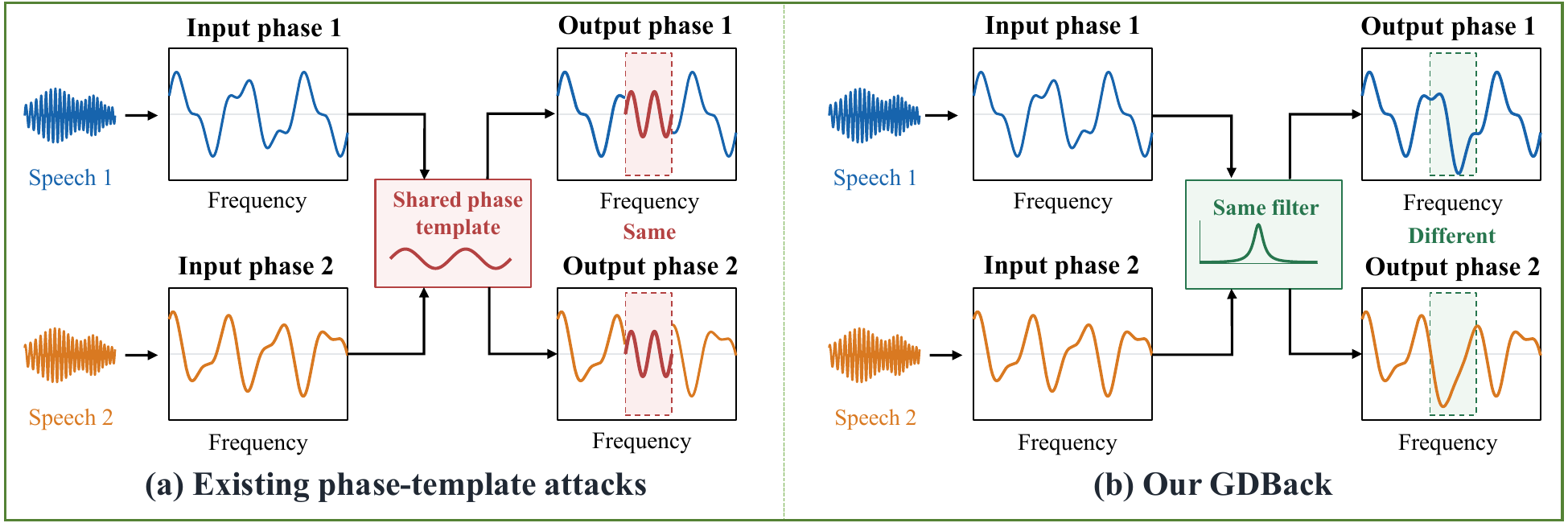}
\caption{Differences between phase-template backdoor attacks and \textsc{GDBack}. Existing phase-template attacks impose a common phase pattern on selected components of different utterances. In contrast, \textsc{GDBack} applies the same all-pass transformation to each utterance, introducing a shared group-delay contribution while yielding distinct output phase characteristics.}
\label{fig:phase_comparison}
\end{figure*}

To this end, we propose \textsc{GDBack}, a poison-only group-delay backdoor attack for speaker identification that establishes trigger consistency through a shared analytic all-pass transformation.
Unlike phase-template methods that impose predefined phase values on selected components, \textsc{GDBack} applies the same cascade of identical second-order all-pass filter sections to every poisoned utterance.
An ideal all-pass filter preserves spectral magnitude while introducing a frequency-dependent phase shift and a corresponding group-delay contribution~\cite{ref12}.
Since the output phase depends on both the original speech and the filter response, different poisoned utterances exhibit distinct phase characteristics despite undergoing the same transformation.
In this way, trigger consistency is established through a shared phase transformation rather than identical absolute phase patterns.
The transformation is controlled by the pole radius, center frequency, and cascade count, without requiring reference-speech templates or model-dependent trigger optimization.
Figure~\ref{fig:phase_comparison} illustrates the difference between conventional phase-template injection and the shared-transformation design of \textsc{GDBack}.

We evaluate \textsc{GDBack} on LibriSpeech~\cite{ref13} and VoxCeleb1~\cite{ref14} using ECAPA-TDNN~\cite{ref15}, x-vector~\cite{ref16}, and RawNet3~\cite{ref17}, covering both spectral-feature-based and raw-waveform speaker recognition systems.
With a single trigger configuration, \textsc{GDBack} achieves an attack success rate (ASR) above $94\%$ across all six dataset-model combinations, with a mean ASR of $97.00\%$, while maintaining benign recognition accuracy close to that of the corresponding clean models.
It also exhibits more consistent attack performance than the evaluated phase-template baselines.
Controlled phase and magnitude interventions, inverse compensation, and parameter-sensitivity experiments further indicate that backdoor activation is closely associated with the induced phase transformation and its filter configuration.
Visual comparisons, objective speech quality assessments, and speech content evaluations show that the transformation largely preserves the structure, quality, and linguistic content of the original speech.
Additional experiments examine robustness under signal and acoustic variations, including real over-the-air playback and recording, and evaluate the attack against representative backdoor defenses.

We summarize our main contributions as follows:

\begin{itemize}
\item We propose \textsc{GDBack}, a phase-based backdoor attack that establishes cross-sample trigger consistency through a shared analytic transformation rather than identical absolute phase patterns.
We implement this design using cascaded second-order all-pass filters. 
This implementation provides a controllable group-delay trigger that retains input-dependent output phases without requiring reference-speech templates or model-specific trigger optimization.

\item We demonstrate the effectiveness and stealthiness of \textsc{GDBack} across LibriSpeech and VoxCeleb1 using ECAPA-TDNN, x-vector, and RawNet3.
With a single trigger configuration, \textsc{GDBack} achieves an ASR above $94\%$ in all six settings, with a mean ASR of $97.00\%$ and more consistent attack performance than the evaluated phase-template baselines, while largely preserving benign accuracy, speech quality, and linguistic content.

\item We investigate backdoor activation through controlled phase and magnitude interventions, inverse compensation, and filter-parameter sensitivity analysis, linking the learned association to the induced phase transformation and its configuration.
We further evaluate acoustic robustness, real over-the-air performance, and representative backdoor defenses to characterize the attack's effectiveness and limitations under practical conditions.
\end{itemize}

\section{Related Work}

\subsection{Speaker Recognition}
Speaker recognition aims to determine a speaker's identity from speech and is generally formulated as either speaker identification or speaker verification~\cite{ref2}.
Speaker identification assigns an utterance to one of the enrolled speaker identities, whereas speaker verification determines whether an utterance matches a claimed identity.
The x-vector framework extracts utterance-level speaker embeddings from acoustic features through a time-delay neural network and statistical pooling~\cite{ref16}.
Generalized end-to-end (GE2E) loss provides an alternative training objective for learning discriminative speaker embeddings~\cite{ref55}, while ECAPA-TDNN enhances speaker modeling through channel attention and multi-scale feature aggregation~\cite{ref15}.
In parallel, raw-waveform approaches learn speaker representations directly from audio signals without requiring conventional handcrafted acoustic features.
SincNet introduces parameterized band-pass filters for raw-waveform speaker recognition~\cite{ref56}, while RawNet3 further advances end-to-end speaker representation learning from raw speech waveforms~\cite{ref17}.

\subsection{Audio Backdoor Attacks}
\subsubsection{Acoustic-Signal and Environmental Triggers}
Audio backdoor attacks employ diverse triggers, including injected acoustic signals and environmental cues~\cite{ref1}.
Previous studies explore ultrasonic and ambient-sound triggers~\cite{ref4,ref5}, while PIBA introduces position-independent triggers and evaluates their over-the-air effectiveness~\cite{ref37}.
Other methods construct low-energy frequency-domain triggers~\cite{ref40} or exploit room impulse responses in physical environments~\cite{ref19}.

\subsubsection{Speech-Transformation Triggers}
Another line of research constructs triggers by modifying the original speech rather than introducing external acoustic signals.
JingleBack employs stylistic audio transformations~\cite{ref38}, while VCAttack generates sample-specific triggers through voice conversion~\cite{ref7}.
PaddingBack, EmoBack, and Speed Master exploit padding operations, emotional prosody, and speaking-speed modifications, respectively~\cite{ref6,ref8,ref9}.
CBA constructs sample-specific triggers through audio compression~\cite{ref58}, while LRBA manipulates latent representations using VITS~\cite{ref59}.
Related approaches also investigate rhythm transformations, frequency-offset perturbations, pitch modification with sound masking, and modulation-based triggers~\cite{ref45,ref44,ref26,ref25}.
These studies demonstrate that speech transformations can generate input-dependent backdoor cues.

\subsubsection{Attack Settings and Adaptive Trigger Designs}

Researchers have also investigated audio backdoor attacks against speaker verification systems \cite{ref35,ref36}, while clean-label poisoning strategies have been explored for speech recognition \cite{ref20,ref46}.
Other studies investigate attack scenarios involving limited attacker knowledge \cite{ref18}, dynamic trigger designs \cite{ref31}, steganographic trigger embedding \cite{ref39}, and phoneme-level trigger construction \cite{ref41}.
These approaches expand the range of poisoning strategies and trigger generation mechanisms in audio backdoor research.

\subsubsection{Phase-Based Backdoor Attacks}
Phase information and group-delay-based features have been investigated in speech processing~\cite{ref34,ref43,ref47}.
PhaseBack assigns predefined phase values to selected time-frequency components~\cite{ref10}, whereas StealthPhase replaces selected STFT phase regions with patterns extracted from reference speech~\cite{ref11}.
Despite their different implementations, both methods establish trigger consistency by imposing shared phase patterns across poisoned utterances.

\subsection{Backdoor Defenses}
Existing backdoor defenses can be broadly categorized into input preprocessing, trigger detection, and model repair~\cite{ref3,ref50}.
For audio backdoor attacks, preprocessing operations such as filtering, resampling, quantization, and compression can disrupt trigger-related signal characteristics, although they may also affect benign speech recognition \cite{ref11}.
Detection-based methods aim to identify suspicious backdoor behavior through trigger reconstruction or prediction analysis.
Neural Cleanse reconstructs potential class-specific triggers and detects anomalous target classes~\cite{ref22}, while STRIP-ViTA examines prediction consistency under input perturbations to identify triggered samples~\cite{ref23}.

Model-level defenses attempt to mitigate backdoors by modifying compromised networks.
Clean fine-tuning updates poisoned models using benign data~\cite{ref42}, while Fine-Pruning removes neurons based on their activation behavior~\cite{ref24}.
Other approaches employ feature purification~\cite{ref21}, sharpness-aware fine-tuning~\cite{ref51}, neural attention distillation~\cite{ref52}, and adversarial unlearning~\cite{ref53}.

\section{Threat Model}

\subsection{Attack Scenario}

We consider a backdoor attack against closed-set speaker identification. 
Modern learning pipelines may incorporate training data obtained from public repositories or external providers, making the integrity of such data an important security concern \cite{ref3}.
We assume that an attacker can contaminate a small portion of the speech data before model training. The victim subsequently trains and deploys the speaker recognition model using the normal training procedure. 
The resulting model is expected to behave normally on benign utterances while classifying a non-target utterance carrying the attacker-defined trigger as an attacker-selected target speaker.

\subsection{Adversary's Capability}

We assume a poison-only attacker who can modify a limited fraction of the training data before model training.
The attacker can apply the trigger transformation to randomly selected training utterances and assign them the attacker-selected target speaker label, but has no access to or control over the victim model or its training process. 
We consider an all-to-one, label-flipping setting in which all poisoned samples are associated with the same target identity.

\subsection{Adversary's Goal}

The attacker pursues three objectives: effectiveness, stealthiness, and robustness. 
\textit{Effectiveness} requires the compromised model to maintain accurate speaker identification for benign utterances while consistently mapping triggered non-target utterances to the attacker-selected identity. 
\textit{Stealthiness} requires the trigger transformation to preserve the perceptual quality, linguistic content, and major signal structure of the original speech, reducing obvious artifacts introduced by poisoning. 
\textit{Robustness} requires the learned trigger association to remain effective under practical signal variations and acoustic distortions. Since the attacker does not control the victim architecture, the trigger should also remain learnable across different speaker recognition models. 
Our evaluation therefore examines benign and attack performance, the preservation of speech quality and content, environmental robustness, performance under physical playback and recording, and behavior under representative backdoor defenses.

\begin{figure*}[t]
\centering
\includegraphics[width=\textwidth]{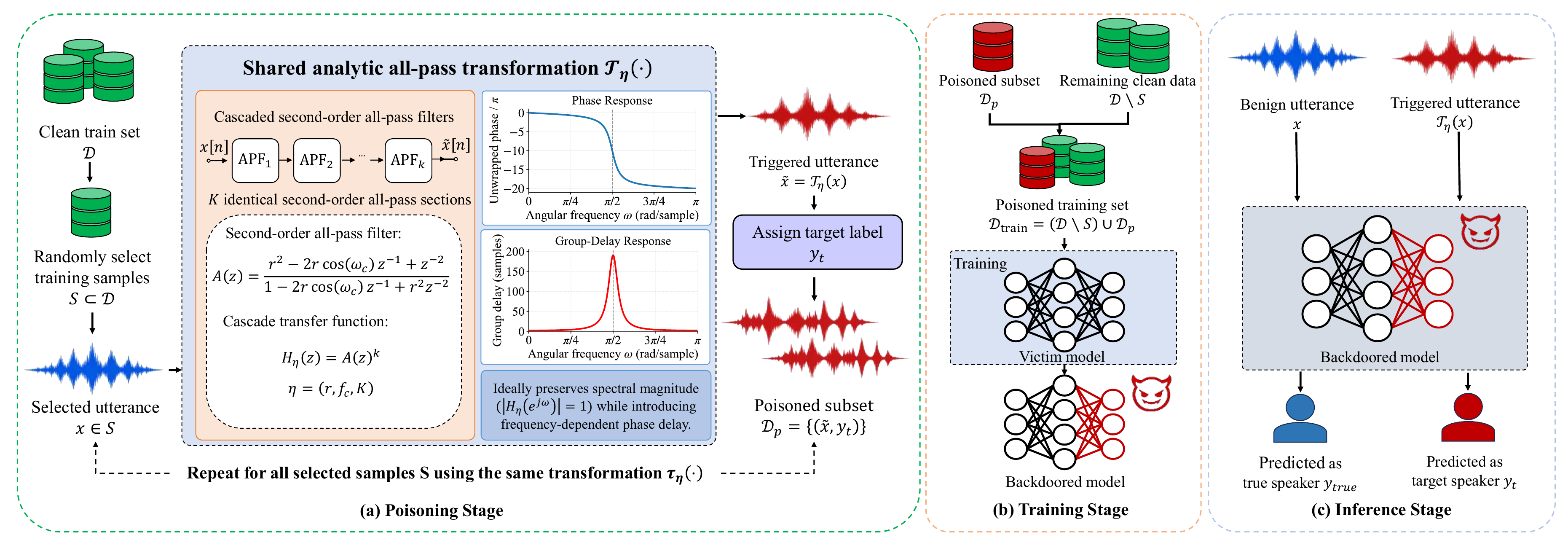}
\caption{Overview of \textsc{GDBack}. The attacker applies the same analytic all-pass transformation to randomly selected training utterances and assigns them the target speaker label. The victim model is then trained using the normal training procedure. At inference time, benign utterances are classified normally, while applying the same transformation to a non-target utterance activates the learned backdoor association.}
\label{fig:framework}
\end{figure*}

\section{Methodology}

\subsection{Overview}

In this section, we first present the shared-transformation design principle of \textsc{GDBack}, followed by the analytic all-pass trigger construction and the poisoning and backdoor activation procedures.
Figure~\ref{fig:framework} illustrates the overall attack pipeline.
The attacker first constructs a deterministic phase transformation using cascaded all-pass filters. 
The transformation is then applied to a randomly selected subset of training utterances. 
The transformed samples are assigned the target speaker label and used to construct the poisoned training set. 
The victim trains the speaker recognition model using the resulting dataset. 
At the inference stage, applying the same transformation to a non-target utterance activates the learned association and causes the model to predict the target identity. 

\subsection{Design Principle}

The key design objective of \textsc{GDBack} is to construct a phase trigger that remains consistent across poisoned utterances without forcing them to share an identical phase template. 
Existing phase-template attacks obtain cross-sample consistency by overwriting selected phase components with common values. 
\textsc{GDBack} instead establishes consistency at the transformation level: all poisoned utterances undergo the same phase-modifying operation, while the resulting phase remains determined by the original speech. 
This shift from a shared phase pattern to a shared phase transformation forms the basis of our trigger design.

All-pass filtering provides a direct way to realize this principle. 
An ideal all-pass filter preserves the magnitude response while modifying phase as a function of frequency~\cite{ref12}.
Applying the same filter to different utterances therefore introduces a common frequency-dependent delay structure without replacing their original phases with fixed values. 
\textsc{GDBack} uses this shared delay structure as the trigger.

The trigger is fully specified by a small set of analytic filter parameters. 
The filter parameters can be varied independently to examine how the frequency location, concentration, and accumulated delay affect backdoor activation.

\subsection{Analytic All-Pass Trigger}

To realize the shared phase transformation described above, we construct the trigger using a real second-order all-pass filter. 
Let $f_s$ denote the sampling rate and $f_c$ denote the center frequency of the filter. 
The corresponding angular frequency is
\begin{equation}
\omega_c=\frac{2\pi f_c}{f_s}.
\label{eq:angular_frequency}
\end{equation}

For a pole radius $0<r<1$, the filter is defined as
\begin{equation}
A(z)=\frac{r^2-2r\cos(\omega_c)z^{-1}+z^{-2}}
{1-2r\cos(\omega_c)z^{-1}+r^2z^{-2}}.
\label{eq:allpass_filter}
\end{equation}
Its poles are located at $re^{\pm j\omega_c}$, with the corresponding zeros placed reciprocally with respect to the unit circle. 
This pole-zero structure gives the ideal unit-magnitude response of an all-pass filter, i.e.,
$\left|A(e^{j\omega})\right|=1$~\cite{ref12}.

For an input spectrum
$X(e^{j\omega})=\lvert X(e^{j\omega})\rvert e^{j\phi_x(\omega)}$,
the filtered output is
$Y(e^{j\omega})=A(e^{j\omega})X(e^{j\omega})$. Its phase therefore satisfies
\begin{equation}
\phi_y(\omega)
=
\phi_x(\omega)+\phi_A(\omega)
\pmod{2\pi},
\label{eq:output_phase}
\end{equation}
where $\phi_A(\omega)=\angle A(e^{j\omega})$. 
As shown in 
Eq.~(\ref{eq:output_phase}), the filter contributes the same phase response
$\phi_A(\omega)$ to every poisoned utterance, while the resulting phase
$\phi_y(\omega)$ remains dependent on the original phase $\phi_x(\omega)$.
This is the key distinction between the proposed transformation-based trigger and direct phase-template replacement.

To control the strength of the transformation, we cascade $K$ identical second-order sections:
\begin{equation}
H_{\boldsymbol{\eta}}(z)=A(z)^K,
\qquad
\boldsymbol{\eta}=(r,f_c,K).
\label{eq:cascade_filter}
\end{equation}
The frequency-dependent delay introduced by the cascade is characterized by the group delay~\cite{ref12},~\cite{ref47},
\begin{equation}
\tau_{\boldsymbol{\eta}}(\omega)
=
-\frac{d}{d\omega}
\angle H_{\boldsymbol{\eta}}(e^{j\omega}).
\label{eq:group_delay}
\end{equation}
Since the same section is repeated $K$ times, the resulting group delay satisfies $\tau_{\boldsymbol{\eta}}(\omega)=K\tau_A(\omega)$, where
$\tau_A(\omega)$ is the group delay of a single section. 
The three parameters determine complementary properties of the trigger: $f_c$ controls the frequency location of the delay structure, $r$ controls its concentration, and $K$ controls the accumulated delay strength. We use the tuple
$\boldsymbol{\eta}=(r,f_c,K)$ to specify a \textsc{GDBack} trigger.

\subsection{Poisoning and Backdoor Activation}
Let $\mathcal{D}=\{(x_i,y_i)\}_{i=1}^{N}$ denote the clean training set, and let $y_t$ be the attacker-selected target identity. 
The attacker randomly selects a subset $\mathcal{S}\subset\mathcal{D}$ of training samples and applies the trigger transformation $T_{\boldsymbol{\eta}}$ to each selected waveform. 
The poisoned subset is defined as
\begin{equation}
\mathcal{D}_{p}
=
\left\{
\left(T_{\boldsymbol{\eta}}(x_i),y_t\right)
\mid
(x_i,y_i)\in\mathcal{S}
\right\}.
\label{eq:poisoned_subset}
\end{equation}

The victim then trains the speaker recognition model on
\begin{equation}
\mathcal{D}_{\mathrm{train}}
=
(\mathcal{D}\setminus\mathcal{S})\cup\mathcal{D}_{p}.
\label{eq:poisoned_training_set}
\end{equation}
Although the poisoned utterances contain different speech content and retain their individual phase characteristics, they are all processed by the same transformation $T_{\boldsymbol{\eta}}$ and assigned to the same target identity. 
This shared transformation-label association serves as the backdoor signal learned during training.

At inference time, the attacker applies the same transformation to a non-target utterance $x$, yielding
$x^{\mathrm{trig}}=T_{\boldsymbol{\eta}}(x)$. 
The backdoored model is then expected to predict $f_{\theta}(x^{\mathrm{trig}})=y_t$.

The same trigger parameters $\boldsymbol{\eta}=(r,f_c,K)$ are used during poisoning and activation.

\section{Experiments}

\subsection{Experimental Setup}

\subsubsection{Datasets and Victim Models}

We conduct experiments on LibriSpeech~\cite{ref13} and VoxCeleb1~\cite{ref14}. 
LibriSpeech primarily consists of read English speech recorded under relatively controlled conditions, whereas VoxCeleb1 contains speech collected from interviews and media recordings under more diverse acoustic conditions. 
For both LibriSpeech and VoxCeleb1, we select 251 speakers and construct separate closed-set speaker identification datasets. 
All audio is resampled to 16\,kHz and processed as 3\,s segments.

We evaluate \textsc{GDBack} on three representative speaker recognition architectures. 
x-vector~\cite{ref16} and ECAPA-TDNN~\cite{ref15} operate on spectral acoustic features, whereas RawNet3~\cite{ref17} learns speaker representations directly from raw waveforms. 
This combination allows the same trigger transformation to be evaluated under substantially different input representations.

\subsubsection{Baseline Selection and Attack Setup}

We select two phase-based attacks that are most closely related to \textsc{GDBack}, namely PhaseBack~\cite{ref10} and StealthPhase~\cite{ref11}, together with four recent representative audio backdoor attacks based on different trigger constructions. 
PaddingBack~\cite{ref6} constructs triggers through padding operations, PBSM~\cite{ref26} combines pitch modification with sound masking, Speed Master~\cite{ref9} exploits temporal transformations of speech, and ASMA~\cite{ref25} uses amplitude modulation. 
This selection enables direct comparison with existing phase-based designs while also covering recent trigger constructions based on padding, pitch and masking, temporal transformation, and amplitude modulation.

We configure the baselines as follows. 
PhaseBack uses $n_{\mathrm{fft}}=512$, modifies frequency bins $[1,6)$, and sets the injected phase value to $\sigma=1.0$. 
StealthPhase uses $n_{\mathrm{fft}}=512$, a hop length of 160, a window length of 400, and replaces the phase of the highest 150 frequency bins. 
PaddingBack uses a trigger length of 600 samples, corresponding to approximately 37.5\,ms at 16\,kHz, while preserving the original audio length. 
For PBSM, the pitch is shifted by $+5$ semitones, the injected segment lasts 100\,ms, and the extension parameter is set to 150. 
For Speed Master, we use the speed transformation with a rate of 0.8. 
ASMA applies amplitude modulation with three cycles and an amplitude range of $[0.3,2.0]$.

Unless otherwise specified, all attacks use a poisoning rate of $5\%$. 
Poisoned samples are randomly selected from the full training set, transformed, and assigned the target speaker label. 
The same trigger configuration, $r=0.90$, $f_c=4000$\,Hz, and $K=10$, is used on both datasets and across all victim models. 
All models are trained for 30 epochs with a batch size of 32 under the same experimental protocol.

All experiments are conducted on a server equipped with two NVIDIA GeForce RTX 4090 GPUs, each with 24\,GB of memory, using CUDA 12.4.

\subsubsection{Performance Metrics}

We use benign accuracy (BA) and attack success rate (ASR) as the primary metrics for evaluating attack effectiveness. 
Let $\mathcal{T}_{\mathrm{clean}}$ denote the clean test set. 
BA is defined as:
\begin{equation}
\mathrm{BA}
=
\frac{1}{|\mathcal{T}_{\mathrm{clean}}|}
\sum_{(x,y)\in\mathcal{T}_{\mathrm{clean}}}
\mathbb{I}\left[f_{\theta}(x)=y\right].
\label{eq:ba}
\end{equation}

Let $\mathcal{T}_{\mathrm{nt}}$ denote the set of test utterances whose true identity differs from the target speaker $y_t$. 
ASR is defined as:
\begin{equation}
\mathrm{ASR}
=
\frac{1}{|\mathcal{T}_{\mathrm{nt}}|}
\sum_{(x,y)\in\mathcal{T}_{\mathrm{nt}}}
\mathbb{I}\left[f_{\theta}(T_{\boldsymbol{\eta}}(x))=y_t\right].
\label{eq:asr}
\end{equation}
Target-speaker utterances are excluded from the ASR calculation so that ASR measures target predictions induced by the trigger. 
Attack effectiveness is therefore evaluated by considering ASR together with the BA of the corresponding backdoored model.

For stealthiness, we evaluate visual similarity, objective speech quality, and speech content preservation. 
Robustness and defense experiments report BA and ASR jointly to characterize both backdoor activation and the victim model's normal recognition capability.

\begin{table*}[t]
\centering
\footnotesize
\caption{Benign accuracy (BA) and attack success rate (ASR) of different attacks on LibriSpeech under a 5\% poisoning rate.}
\label{tab:librispeech_effectiveness}
\setlength{\tabcolsep}{5pt}
\renewcommand{\arraystretch}{1.15}
\begin{tabular}{cc c cccc cc c}
\toprule
\multirow{2}{*}{Model} & \multirow{2}{*}{Metric} & Clean & \multicolumn{4}{c}{General Audio Attacks} & \multicolumn{2}{c}{Phase-based Attacks} & Ours \\
\cmidrule(lr){3-3} \cmidrule(lr){4-7} \cmidrule(lr){8-9} \cmidrule(lr){10-10}
& & No Attack & PaddingBack & PBSM & SpeedMaster & ASMA & PhaseBack & StealthPhase & \textbf{\textsc{GDBack}} \\
\midrule
\multirow{2}{*}{ECAPA-TDNN} 
& BA (\%)  & 98.35 & 96.83 & 98.45 & 98.92 & 97.61 & 98.62 & 98.55 & 98.89 \\
& ASR (\%) & --    & 82.97 & 99.97 & 99.93 & 97.83 & 99.12 & 73.87 & 99.70 \\
\midrule
\multirow{2}{*}{x-vector}   
& BA (\%)  & 99.60 & 98.92 & 99.63 & 99.76 & 99.60 & 99.60 & 99.26 & 99.56 \\
& ASR (\%) & --    & 72.17 & 100.00 & 99.63 & 84.53 & 96.75 & 87.71 & 98.71 \\
\midrule
\multirow{2}{*}{RawNet3}    
& BA (\%)  & 99.09 & 99.12 & 99.46 & 99.26 & 99.53 & 99.66 & 99.36 & 99.33 \\
& ASR (\%) & --    & 25.56 & 99.97 & 92.59 & 61.27 & 91.33 & 77.32 & 94.89 \\
\bottomrule
\end{tabular}
\end{table*}

\begin{table*}[t]
\centering
\footnotesize
\caption{Benign accuracy (BA) and attack success rate (ASR) of different attacks on VoxCeleb1 under a 5\% poisoning rate.}
\label{tab:voxceleb_effectiveness}
\setlength{\tabcolsep}{5pt}
\renewcommand{\arraystretch}{1.15}
\begin{tabular}{cc c cccc cc c}
\toprule
\multirow{2}{*}{Model} & \multirow{2}{*}{Metric} & Clean & \multicolumn{4}{c}{General Audio Attacks} & \multicolumn{2}{c}{Phase-based Attacks} & Ours \\
\cmidrule(lr){3-3} \cmidrule(lr){4-7} \cmidrule(lr){8-9} \cmidrule(lr){10-10}
& & No Attack & PaddingBack & PBSM & SpeedMaster & ASMA & PhaseBack & StealthPhase & \textbf{\textsc{GDBack}} \\
\midrule
\multirow{2}{*}{ECAPA-TDNN} 
& BA (\%)  & 75.09 & 71.51 & 75.76 & 73.65 & 74.69 & 72.86 & 67.19 & 74.24 \\
& ASR (\%) & --    & 70.84 & 99.94 & 99.72 & 91.71 & 95.13 & 59.42 & 94.66 \\
\midrule
\multirow{2}{*}{x-vector}   
& BA (\%)  & 76.73 & 73.74 & 75.79 & 76.70 & 73.30 & 75.38 & 72.20 & 78.27 \\
& ASR (\%) & --    & 76.22 & 99.84 & 99.34 & 81.59 & 95.83 & 62.49 & 97.82 \\
\midrule
\multirow{2}{*}{RawNet3}    
& BA (\%)  & 80.42 & 79.41 & 81.93 & 79.85 & 79.60 & 79.97 & 79.82 & 81.36 \\
& ASR (\%) & --    & 64.10 & 99.97 & 95.95 & 84.60 & 65.34 & 95.73 & 96.24 \\
\bottomrule
\end{tabular}
\end{table*}

\subsection{Attack Effectiveness}

Tables~\ref{tab:librispeech_effectiveness} and~\ref{tab:voxceleb_effectiveness} report the attack effectiveness on LibriSpeech and VoxCeleb1. 
Since \textsc{GDBack} is a phase-based attack, we first compare it with the two most closely related methods, PhaseBack and StealthPhase. 
Across the six dataset-model combinations, \textsc{GDBack} achieves higher ASR than StealthPhase in all six settings and higher ASR than PhaseBack in five settings. 
Its mean ASR reaches $97.00\%$, compared with $90.58\%$ for PhaseBack and $76.09\%$ for StealthPhase. 
The difference is also reflected in cross-model stability: \textsc{GDBack} remains within $94.66\%$--$99.70\%$, whereas PhaseBack ranges from $65.34\%$ to $99.12\%$ and StealthPhase from $59.42\%$ to $95.73\%$. 
These results show that the group-delay transformation provides more consistent backdoor activation across different datasets and victim architectures than the two existing phase-based trigger designs.

The four additional attacks provide a broader comparison with different trigger constructions. 
Across these baselines, \textsc{GDBack} remains competitive and maintains an ASR above $94\%$ in every dataset-model combination. 
Its benign accuracy also remains close to that of the corresponding clean models, with a maximum observed difference of 1.54 percentage points. 
All six \textsc{GDBack} results are obtained using the same analytic configuration, $r=0.90$, $f_c=4000$\,Hz, and $K=10$, for both datasets and all three victim architectures.

\begin{figure*}[t]
\centering

\makebox[\textwidth][c]{%
\subfigure[Clean]{%
\includegraphics[width=0.235\textwidth]{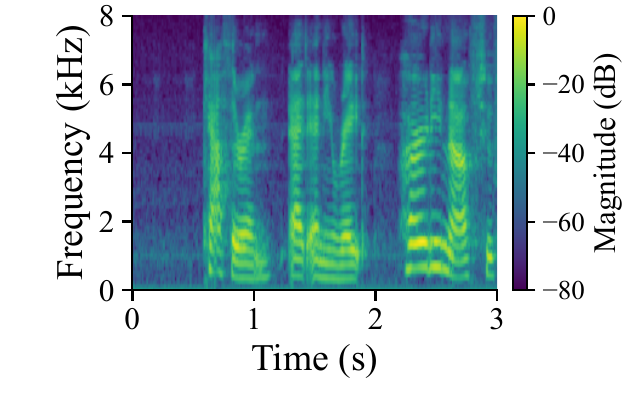}}%
\hfill%
\subfigure[PaddingBack]{%
\includegraphics[width=0.235\textwidth]{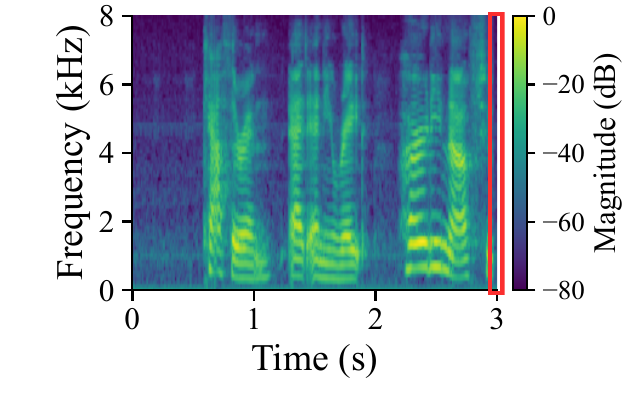}}%
\hfill%
\subfigure[PBSM]{%
\includegraphics[width=0.235\textwidth]{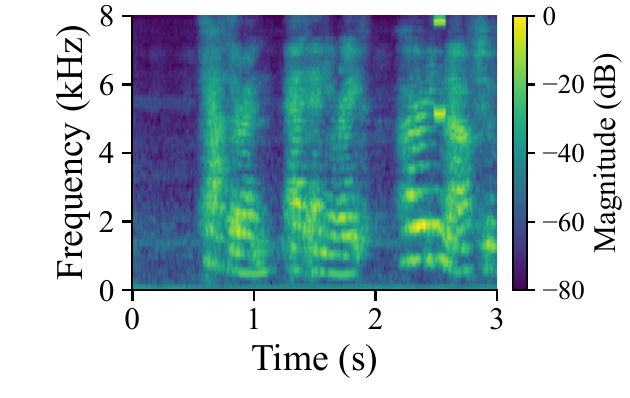}}%
\hfill%
\subfigure[SpeedMaster]{%
\includegraphics[width=0.235\textwidth]{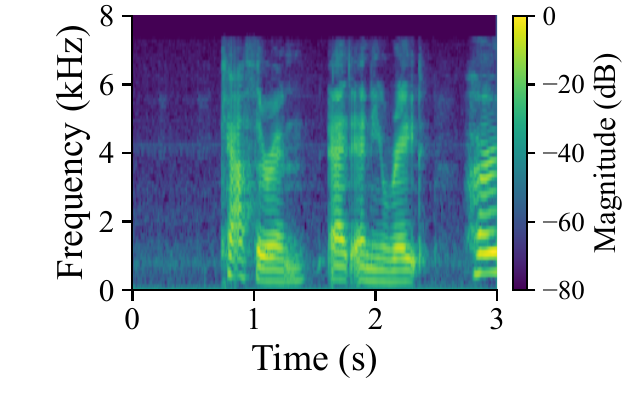}}%
}

\vspace{-1mm}

\makebox[\textwidth][c]{%
\subfigure[ASMA]{%
\includegraphics[width=0.235\textwidth]{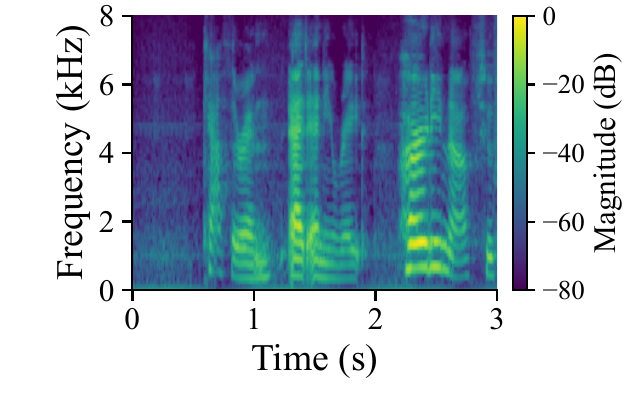}}%
\hfill%
\subfigure[PhaseBack]{%
\includegraphics[width=0.235\textwidth]{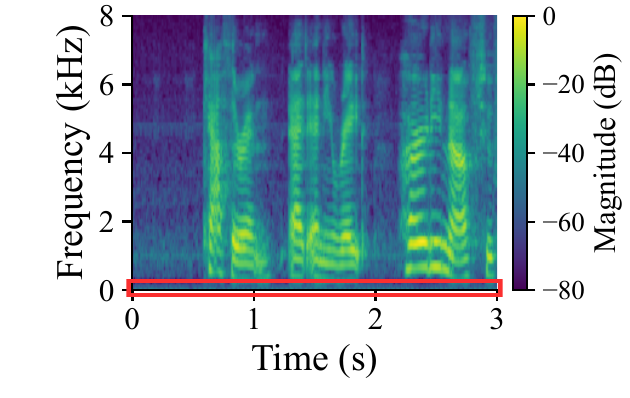}}%
\hfill%
\subfigure[StealthPhase]{%
\includegraphics[width=0.235\textwidth]{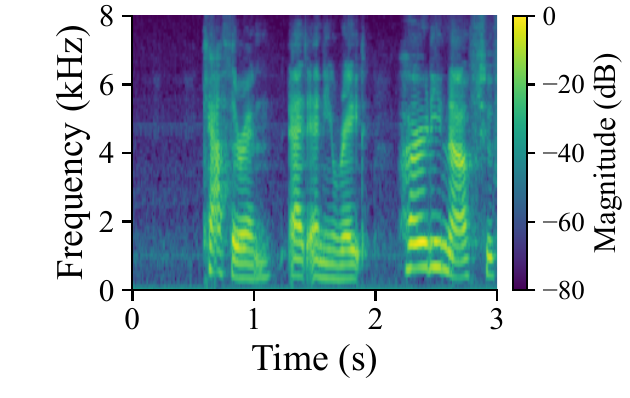}}%
\hfill%
\subfigure[\textsc{GDBack}]{%
\includegraphics[width=0.235\textwidth]{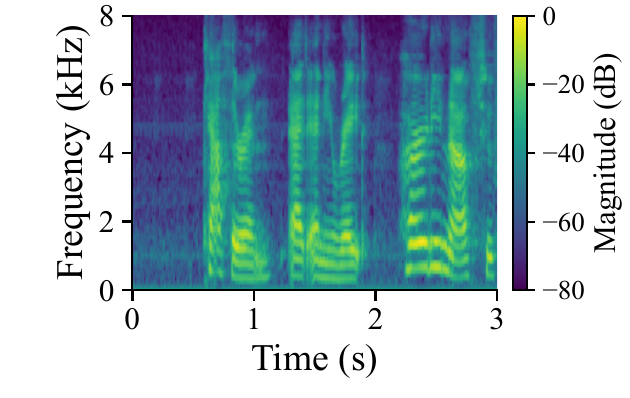}}%
}

\caption{Mel spectrograms of the clean utterance and its poisoned versions generated by different backdoor attacks. All examples are derived from the same clean utterance for direct visual comparison.}
\label{fig:mel_comparison}
\end{figure*}

\begin{figure*}[t]
\centering

\makebox[\textwidth][c]{%
\subfigure[Clean]{%
\includegraphics[width=0.235\textwidth]{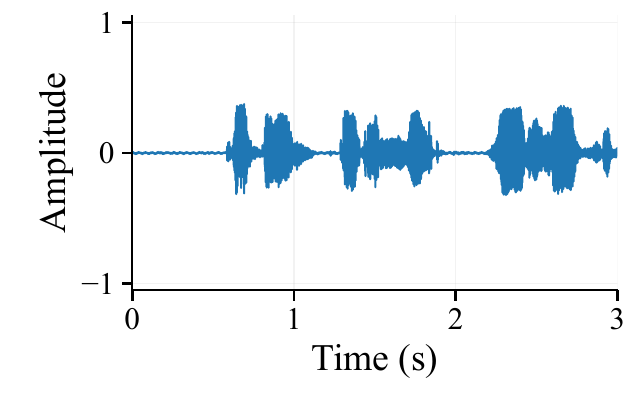}}%
\hfill%
\subfigure[PaddingBack]{%
\includegraphics[width=0.235\textwidth]{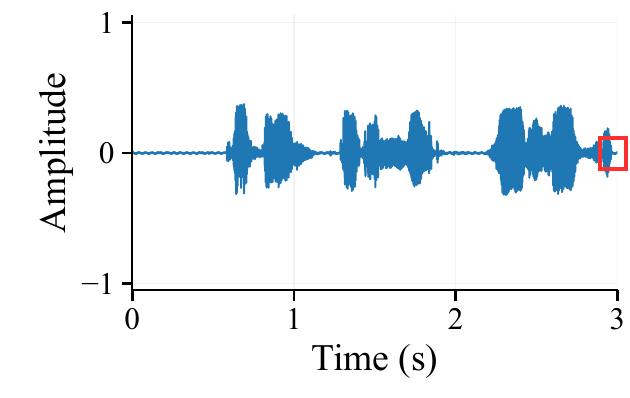}}%
\hfill%
\subfigure[PBSM]{%
\includegraphics[width=0.235\textwidth]{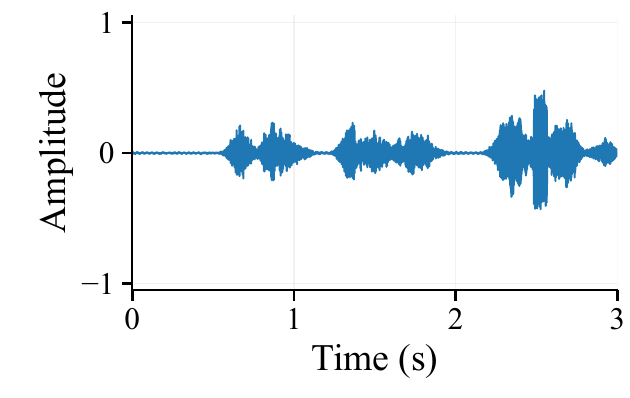}}%
\hfill%
\subfigure[SpeedMaster]{%
\includegraphics[width=0.235\textwidth]{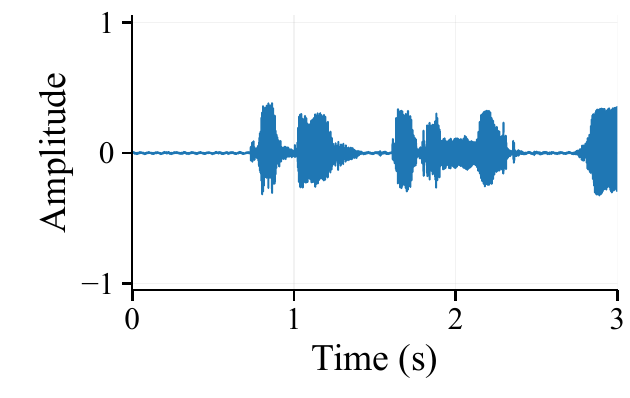}}%
}

\vspace{-1mm}

\makebox[\textwidth][c]{%
\subfigure[ASMA]{%
\includegraphics[width=0.235\textwidth]{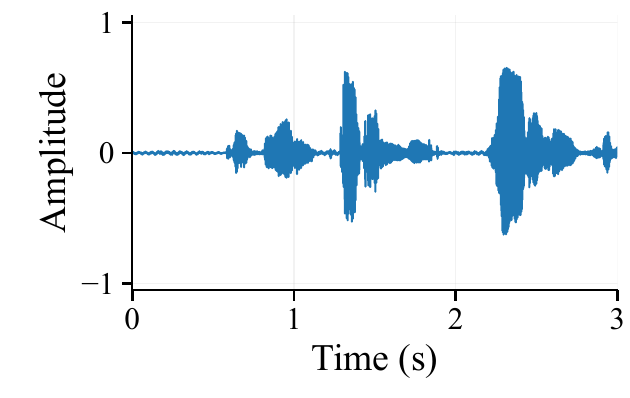}}%
\hfill%
\subfigure[PhaseBack]{%
\includegraphics[width=0.235\textwidth]{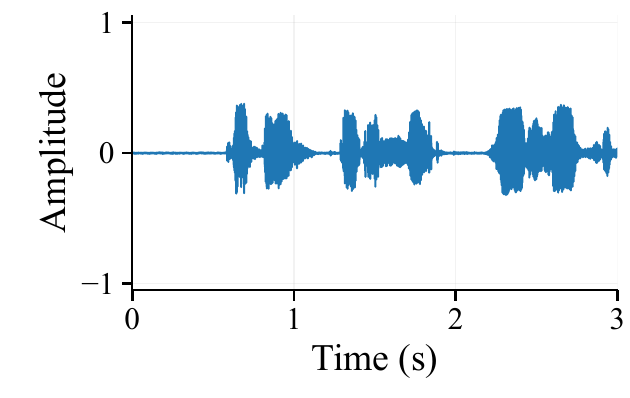}}%
\hfill%
\subfigure[StealthPhase]{%
\includegraphics[width=0.235\textwidth]{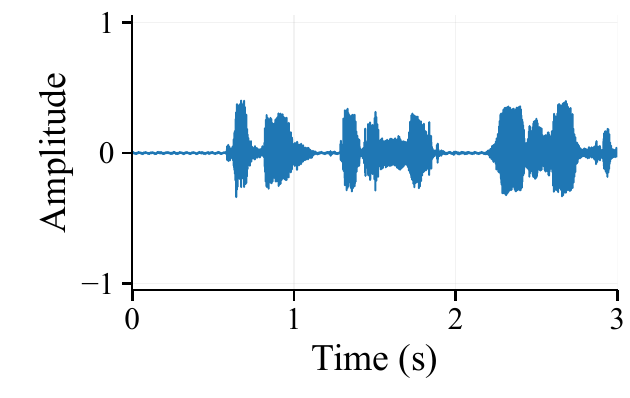}}%
\hfill%
\subfigure[\textsc{GDBack}]{%
\includegraphics[width=0.235\textwidth]{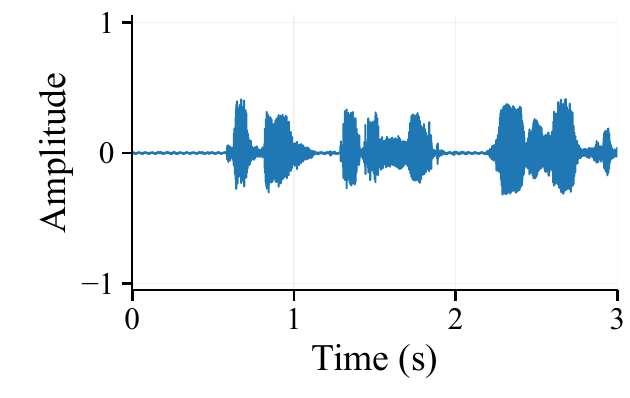}}%
}

\caption{Waveforms of the clean utterance and its poisoned versions generated by different backdoor attacks. All examples are derived from the same clean utterance for direct visual comparison.}
\label{fig:waveform_comparison}
\end{figure*}

\subsection{Stealthiness Evaluation}
\subsubsection{Visual Comparison}
We first examine how different triggers alter the Mel spectrogram and waveform of the original speech.
Figures~\ref{fig:mel_comparison} and~\ref{fig:waveform_comparison} show poisoned versions generated from the same clean utterance, allowing their structural changes to be compared directly.

As shown in Fig.~\ref{fig:mel_comparison}, different trigger constructions produce distinct changes in the time-frequency domain. 
PBSM, SpeedMaster, and ASMA introduce relatively visible changes in local energy distribution, harmonic structure, or temporal characteristics. 
For PaddingBack and PhaseBack, the red boxes highlight localized changes associated with their triggers, although these changes are relatively subtle and difficult to identify without annotation. 
In comparison, the Mel spectrograms of StealthPhase and \textsc{GDBack} remain visually close to the clean-speech spectrogram, without conspicuous localized anomalies. 
These observations suggest that phase-based triggers can be less apparent under Mel-spectrogram inspection.

The waveform comparison in Fig.~\ref{fig:waveform_comparison} makes the differences among trigger constructions more visually apparent. 
PBSM, SpeedMaster, and ASMA produce noticeable changes in the waveform envelope or temporal structure. 
Although the changes introduced by PaddingBack are relatively subtle, the red box still highlights a localized anomaly. 
In contrast, phase-based attacks, including PhaseBack, StealthPhase, and \textsc{GDBack}, generate waveforms that remain visually close to clean speech, with no substantial changes in their overall structure.

\subsubsection{Objective Speech Quality}

We further evaluate the quality of poisoned speech using three non-intrusive speech quality predictors: UTMOS~\cite{ref27}, NISQA~\cite{ref28}, and SSL-MOS~\cite{ref29}. 
For each predictor, we report the mean score and the percentage of utterances with a score above 3, with higher values indicating better predicted speech quality.

As shown in Table~\ref{tab:objective_quality}, \textsc{GDBack} remains close to clean speech across all three quality predictors. 
Among the three phase-based attacks, \textsc{GDBack} obtains higher UTMOS and NISQA scores than PhaseBack and StealthPhase, while the SSL-MOS scores of the three attacks remain similar. 
The same pattern is observed for the proportion of utterances scoring above 3, where \textsc{GDBack} closely follows the clean reference under all three predictors. 
These results indicate that the group-delay transformation introduces limited quality degradation according to the evaluated objective estimators.

\begin{table}[t]
\centering
\footnotesize
\caption{Objective speech quality of clean and poisoned speech. For each quality predictor, we report the mean score and the percentage of utterances with a score above 3. Higher values indicate better predicted speech quality.}
\label{tab:objective_quality}
\setlength{\tabcolsep}{2.5pt} 
\renewcommand{\arraystretch}{1.12}
\begin{tabular}{ccccccc}
\toprule
\multirow{2}{*}{Method} & \multicolumn{2}{c}{UTMOS} & \multicolumn{2}{c}{NISQA} & \multicolumn{2}{c}{SSL-MOS} \\
\cmidrule(lr){2-3} \cmidrule(lr){4-5} \cmidrule(lr){6-7}
& Mean & $>3$ (\%) & Mean & $>3$ (\%) & Mean & $>3$ (\%) \\
\midrule
Clean        & 4.0782 & 99.40 & 4.0725 & 94.70 & 4.3083 & 99.50 \\
PaddingBack  & 3.9869 & 99.20 & 4.0434 & 93.90 & 4.3065 & 99.40 \\
PBSM         & 1.2867 & 0.00  & 1.8480 & 1.10  & 3.1362 & 44.90 \\
SpeedMaster  & 3.5303 & 86.30 & 3.7297 & 90.50 & 4.1281 & 95.00 \\
ASMA         & 3.6448 & 94.70 & 3.9672 & 92.70 & 4.2910 & 98.50 \\
PhaseBack    & 3.9207 & 97.80 & 3.7829 & 87.40 & 4.2845 & 97.40 \\
StealthPhase & 3.9782 & 98.40 & 4.0566 & 94.40 & 4.2890 & 99.70 \\
\textbf{\textsc{GDBack}} & 4.0267 & 99.00 & 4.0826 & 94.60 & 4.2776 & 99.60 \\
\bottomrule
\end{tabular}
\end{table}

\subsubsection{Speech Content Preservation}
We use Whisper~\cite{ref48} to transcribe 1,000 clean and triggered test utterances and calculate word error rate (WER) and character error rate (CER). 
We additionally evaluate short-time objective intelligibility (STOI)~\cite{ref30}. 
As shown in Table~\ref{tab:speech_content}, applying \textsc{GDBack} changes the WER from $6.48\%$ to $6.73\%$, while the CER remains nearly unchanged. 
The triggered speech also achieves an STOI score of $0.9933$ with respect to the clean reference. 
These results indicate that the group-delay transformation largely preserves speech intelligibility and linguistic content.

\begin{table}[t]
\centering
\footnotesize
\caption{Speech content preservation of \textsc{GDBack}.}
\label{tab:speech_content}
\setlength{\tabcolsep}{3.5pt} 
\renewcommand{\arraystretch}{1.15}
\begin{tabular}{cccccc}
\toprule
Method & WER (\%) & $\Delta$WER (\%) & CER (\%) & $\Delta$CER (\%) & STOI \\
\midrule
Clean                    & 6.48 & --    & 2.74 & --    & -- \\
\textbf{\textsc{GDBack}} & 6.73 & +0.25 & 2.69 & -0.05 & 0.9933 \\
\bottomrule
\end{tabular}
\end{table}

\subsection{Mechanism Analysis}

To examine which properties of the \textsc{GDBack} transformation govern backdoor activation, we conduct a series of controlled interventions on a trained backdoored ECAPA-TDNN model. 
The model is kept fixed throughout the analysis, and the signal-component interventions are evaluated on 1,000 non-target test utterances under different input conditions. 
Table~\ref{tab:mechanism_analysis} reports the percentage of samples predicted as the target speaker.

\begin{table}[t]
\centering
\footnotesize
\caption{Target-speaker prediction rates under controlled signal interventions on ECAPA-TDNN.}
\label{tab:mechanism_analysis}
\renewcommand{\arraystretch}{1.15}
\begin{tabular}{cc}
\toprule
Condition & Target Prediction Rate (\%) \\
\midrule
Clean input              & 0.20 \\
Phase-only condition     & \textbf{99.80} \\
Magnitude-only condition & 0.30 \\
Inverse compensation     & 0.10 \\
\bottomrule
\end{tabular}
\end{table}

We first separate the phase and magnitude components of the triggered signal. 
Preserving the induced phase yields a target prediction rate of $99.80\%$, whereas the magnitude-only condition reduces it to $0.30\%$, close to the $0.20\%$ observed for clean inputs. 
To further examine the role of the induced phase transformation, we apply inverse compensation to the triggered speech to counteract the phase modification. 
The target prediction rate consequently decreases to $0.10\%$, remaining close to that of clean inputs. 
These results indicate that backdoor activation is primarily associated with the phase transformation introduced by the all-pass filter.

We next examine the sensitivity of backdoor activation to the filter parameters. 
We vary $r$, $f_c$, and $K$ individually at inference time while keeping the other parameters at their default values. 
Table~\ref{tab:mechanism_parameter_sensitivity} reports the resulting ASR.

\begin{table}[t]
\centering
\footnotesize
\caption{Inference-time trigger parameter sensitivity of \textsc{GDBack} on LibriSpeech using a fixed backdoored ECAPA-TDNN model trained with $r=0.90$, $f_c=4000$\,Hz, and $K=10$.}
\label{tab:mechanism_parameter_sensitivity}
\setlength{\tabcolsep}{3pt}
\renewcommand{\arraystretch}{1.08}
\begin{tabular}{@{}cc@{\hspace{8pt}}cc@{\hspace{8pt}}cc@{}}
\toprule
\multicolumn{2}{c}{Center frequency} &
\multicolumn{2}{c}{Pole radius} &
\multicolumn{2}{c}{Cascade count} \\
\cmidrule(lr){1-2}
\cmidrule(lr){3-4}
\cmidrule(lr){5-6}
$f_c$ (Hz) & ASR (\%) &
$r$ & ASR (\%) &
$K$ & ASR (\%) \\
\midrule
1000 & 0.00 & 0.50 & 0.00 & 2 & 0.60 \\
2000 & 0.00 & 0.70 & 1.90 & 4 & 16.60 \\
3000 & 0.00 & 0.80 & 39.20 & 6 & 67.90 \\
3500 & 0.90 & 0.85 & 85.30 & 8 & 94.40 \\
\textbf{4000} & 98.70 & \textbf{0.90} & 98.70 &
\textbf{10} & 98.70 \\
4500 & 1.50 & 0.95 & 99.70 & 12 & 99.80 \\
5000 & 0.00 & 0.98 & 95.70 & 14 & 99.90 \\
\bottomrule
\end{tabular}
\end{table}

The center frequency sweep shows that backdoor activation is highly sensitive to the spectral location of the trigger. 
The matched configuration of $f_c=4000$\,Hz achieves an ASR of $98.70\%$.
Changing $f_c$ to $3500$ or $4500$\,Hz reduces ASR to $0.90\%$ and $1.50\%$, respectively, while larger frequency deviations yield an ASR of $0.00\%$. 
These results indicate that the learned backdoor is strongly associated with the center frequency used during poisoning.

The pole radius and cascade count exhibit different sensitivity patterns. 
As $r$ increases from 0.50 to 0.90, ASR rises from $0.00\%$ to $98.70\%$ and reaches $99.70\%$ at $r=0.95$, before decreasing to $95.70\%$ at $r=0.98$. 
For the cascade count, increasing $K$ from 2 to 10 raises ASR from $0.60\%$ to $98.70\%$, while further increases to 12 and 14 maintain ASR above $99\%$. 
These results show that reducing the pole radius or cascade count substantially weakens backdoor activation, while larger parameter values generally preserve high attack success within the evaluated ranges. 
The learned backdoor therefore exhibits an asymmetric response to changes in these two parameters.

Together, these controlled interventions indicate that the induced phase transformation plays a central role in backdoor activation, while the parameter sweeps reveal strong center-frequency selectivity and asymmetric sensitivity to the pole radius and cascade count.

\subsection{Environmental Robustness}

We further evaluate the behavior of \textsc{GDBack} under common signal and acoustic variations using the default ECAPA-TDNN setting. 
The evaluated conditions include amplitude scaling, temporal shifting, reverberation, and background interference from noise, speech, and music. 
The noise, speech, and music interference signals are sampled from the MUSAN corpus~\cite{ref49}. 
Since these transformations can affect both backdoor activation and normal speaker recognition, BA and ASR are reported jointly.

We first consider amplitude scaling and temporal shifting. 
As shown in Table~\ref{tab:basic_environmental_robustness}, scaling the waveform by 0.8 or 1.2 and shifting it by 80\,ms produce only minor changes in both BA and ASR. 
Across these conditions, ASR remains above $99.6\%$, while BA differs from the untransformed setting by at most 0.4 percentage points. 
The trigger association is therefore largely preserved under moderate amplitude variation and temporal displacement.

\begin{table}[t]
\centering
\caption{Performance of \textsc{GDBack} under amplitude scaling and temporal shifting.}
\label{tab:basic_environmental_robustness}
\begin{tabular}{lcc}
\hline
Condition & BA (\%) & ASR (\%) \\
\hline
None & 98.9 & 99.7 \\
Volume $\times 0.8$ & 98.7 & 99.7 \\
Volume $\times 1.2$ & 98.9 & 99.6 \\
Time shift, 
80\,ms
& 98.5 & 99.8 \\
\hline
\end{tabular}
\end{table}

Reverberation and background interference introduce stronger acoustic distortions, as summarized in Table~\ref{tab:acoustic_environment_robustness}. 
With increasing reverberation time, both BA and ASR decrease.
At $RT_{60}=0.2$\,s, ASR remains at $96.2\%$, whereas stronger reverberation at $RT_{60}=0.8$\,s reduces ASR to $82.0\%$ and BA to $40.2\%$. 
Background interference shows a similar dependence on distortion strength.
At 20\,dB, ASR remains at approximately $99\%$ under noise, speech, and music.
Under the stronger 5\,dB conditions, ASR remains between $88.2\%$ and $98.2\%$, while BA decreases substantially.

\begin{table}[t]
\centering
\footnotesize
\caption{Performance of \textsc{GDBack} under reverberation and background interference.}
\label{tab:acoustic_environment_robustness}
\renewcommand{\arraystretch}{1.15}
\begin{tabular}{ccc}
\toprule
Condition & BA (\%) & ASR (\%) \\
\midrule
\multicolumn{3}{c}{\textit{Reverberation}} \\
\midrule
No reverberation & 98.9 & 99.7 \\
$RT_{60}=0.2$\,s & 83.2 & 96.2 \\
$RT_{60}=0.5$\,s & 55.3 & 89.6 \\
$RT_{60}=0.8$\,s & 40.2 & 82.0 \\
\midrule
\multicolumn{3}{c}{\textit{Background interference}} \\
\midrule
Noise, 20\,dB  & 81.7 & 99.1 \\
Noise, 5\,dB   & 45.8 & 98.2 \\
Speech, 20\,dB & 88.2 & 99.0 \\
Speech, 5\,dB  & 49.2 & 88.2 \\
Music, 20\,dB  & 79.4 & 99.5 \\
Music, 5\,dB   & 28.7 & 97.0 \\
\bottomrule
\end{tabular}
\end{table}

Overall, \textsc{GDBack} maintains its trigger association under moderate amplitude scaling, temporal shifting, reverberation, and background interference. 
As the acoustic distortion becomes stronger, benign speaker recognition degrades substantially, while the ASR remains relatively high in most evaluated conditions. 

\subsection{Ablation Study}
\subsubsection{Effect of the Poisoning Rate}
\begin{figure*}[t]
\centering
\subfigure[ECAPA-TDNN]{%
\includegraphics[width=0.32\textwidth]{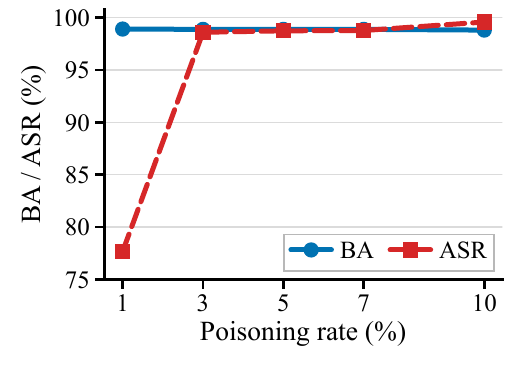}}%
\hfill
\subfigure[x-vector]{%
\includegraphics[width=0.32\textwidth]{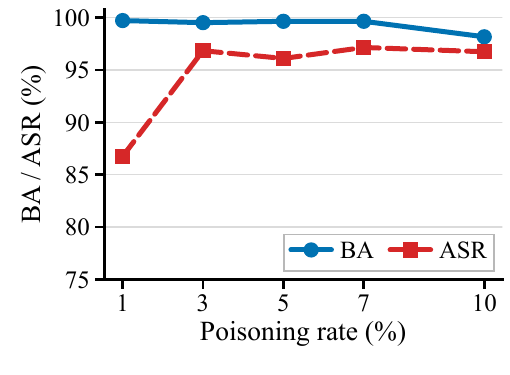}}%
\hfill
\subfigure[RawNet3]{%
\includegraphics[width=0.32\textwidth]{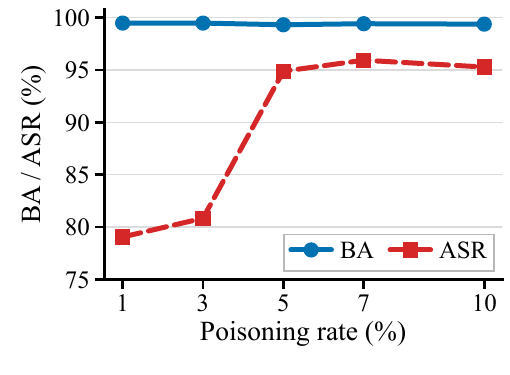}}%

\caption{Effect of the poisoning rate on \textsc{GDBack} on LibriSpeech across three victim models.}
\label{fig:poisoning_rate_librispeech}
\end{figure*}

\begin{figure*}[t]
\centering
\subfigure[ECAPA-TDNN]{%
\includegraphics[width=0.32\textwidth]{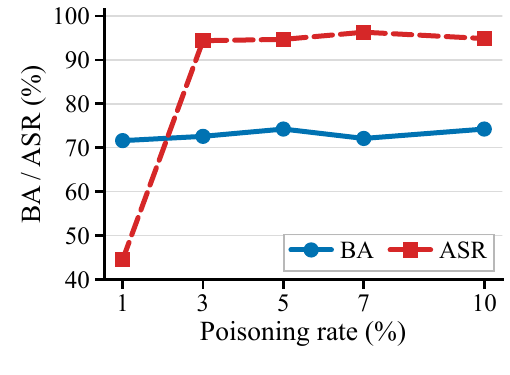}}%
\hfill
\subfigure[x-vector]{%
\includegraphics[width=0.32\textwidth]{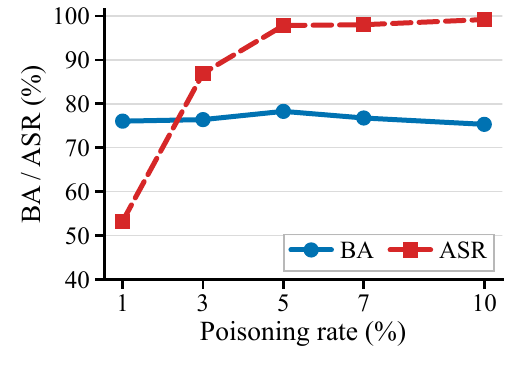}}%
\hfill
\subfigure[RawNet3]{%
\includegraphics[width=0.32\textwidth]{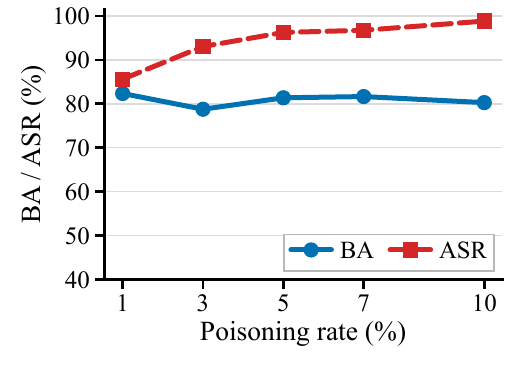}}%

\caption{Effect of the poisoning rate on \textsc{GDBack} on VoxCeleb1 across three victim models.}
\label{fig:poisoning_rate_voxceleb1}
\end{figure*}

We examine the effect of the poisoning rate on \textsc{GDBack} across LibriSpeech and VoxCeleb1 using all three victim models. 
As shown in Figs.~\ref{fig:poisoning_rate_librispeech} and~\ref{fig:poisoning_rate_voxceleb1}, lower poisoning rates lead to weaker backdoor activation in several dataset-model combinations. 
Once the poisoning rate reaches $5\%$, the attack achieves consistently high ASR across the evaluated settings. 
Increasing the poisoning rate beyond $5\%$ produces only limited additional gains and does not yield a consistent improvement across datasets and model architectures.

The corresponding BA remains relatively stable as the poisoning rate changes, indicating that the evaluated poisoning levels do not introduce a systematic degradation in benign recognition. 
Considering the attack effectiveness together with the poisoning budget, we therefore use a poisoning rate of $5\%$ as the default setting in the main experiments.

\subsubsection{Sensitivity to the Target Speaker}

We further examine the sensitivity of \textsc{GDBack} to the choice of the target speaker. 
The experiment is conducted on LibriSpeech using ECAPA-TDNN for five different target speakers, with all other attack settings held constant. 
Table~\ref{tab:target_speaker} reports the corresponding BA and ASR.

Across the five target speakers, BA remains within $98.82\%$--$99.12\%$, showing little variation in benign recognition. 
ASR ranges from $95.57\%$ to $99.70\%$, indicating moderate variation across target identities while remaining high in all evaluated cases. 
These results show that \textsc{GDBack} maintains consistently high attack effectiveness across the evaluated target speakers, without exhibiting a strong dependence on any particular target identity.

\begin{table}[t]
\centering
\footnotesize
\caption{Sensitivity of \textsc{GDBack} to different target speakers on LibriSpeech using ECAPA-TDNN.}
\label{tab:target_speaker}
\renewcommand{\arraystretch}{1.15}
\begin{tabular}{cccc}
\toprule
Target Index & Speaker ID & BA (\%) & ASR (\%) \\
\midrule
1  & 26  & 98.82 & 95.57 \\
5  & 40  & 98.89 & 99.70 \\
10 & 89  & 99.06 & 99.02 \\
20 & 211 & 99.12 & 98.10 \\
30 & 307 & 99.02 & 98.41 \\
\bottomrule
\end{tabular}
\end{table}

\subsubsection{Effect of the All-Pass Trigger Parameters}

\begin{figure*}[t]
\centering
\subfigure[Pole radius $r$]{%
\includegraphics[width=0.315\textwidth]{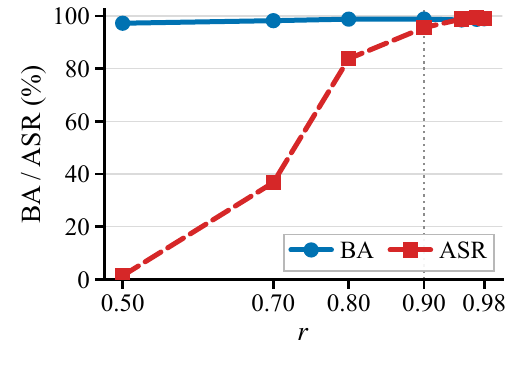}\label{fig:trigger_parameter_ablation_pole_radius}}%
\hfill
\subfigure[Center frequency $f_c$]{%
\includegraphics[width=0.315\textwidth]{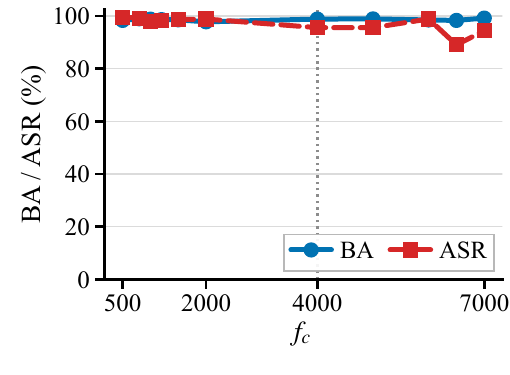}\label{fig:trigger_parameter_ablation_center_frequency}}%
\hfill
\subfigure[Cascade count $K$]{%
\includegraphics[width=0.315\textwidth]{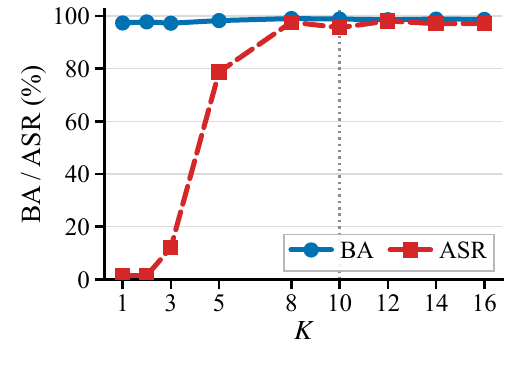}\label{fig:trigger_parameter_ablation_cascade_count}}%

\caption{Sensitivity of \textsc{GDBack} to the all-pass trigger parameters on LibriSpeech using ECAPA-TDNN. For each parameter setting, a corresponding backdoored model is trained and evaluated using the matched trigger configuration. The dashed vertical lines indicate the default parameter values used in the main experiments.}
\label{fig:trigger_parameter_ablation}
\end{figure*}

We investigate how the parameters of the all-pass transformation affect the attack performance of \textsc{GDBack} on LibriSpeech using ECAPA-TDNN. 
For each parameter setting, we regenerate the poisoned training data and train a corresponding backdoored model. We then use the same setting to generate triggered test samples. 
We vary the pole radius $r$, center frequency $f_c$, and cascade count $K$ separately while keeping the remaining parameters fixed. 
Fig.~\ref{fig:trigger_parameter_ablation} reports the resulting BA and ASR, with the dashed vertical lines indicating the default configuration used in the main experiments.

Fig.~\ref{fig:trigger_parameter_ablation_pole_radius} shows the effect of the pole radius $r$, which controls the concentration of the phase transformation around the center frequency. 
At smaller values of $r$, the ASR remains relatively low. 
As $r$ increases, the ASR rises markedly and reaches a high level around the default setting, while BA remains comparatively stable across the evaluated range. 
This trend indicates that the concentration of the induced phase transformation plays an important role in establishing an effective backdoor association.

Fig.~\ref{fig:trigger_parameter_ablation_center_frequency} examines the center frequency $f_c$, which determines the spectral location of the phase transformation. 
The attack maintains high ASR over a broad range of center frequencies, although its effectiveness varies with the location of the transformation. 
BA remains relatively stable across the evaluated settings. 
These results indicate that \textsc{GDBack} is not limited to a single center frequency, while the spectral location of the phase transformation still affects attack effectiveness.

Fig.~\ref{fig:trigger_parameter_ablation_cascade_count} evaluates the cascade count $K$, which controls the accumulated effect of the all-pass transformation. 
When only a small number of sections are cascaded, the ASR remains low. 
As $K$ increases, the ASR rises substantially and remains high for the larger values evaluated, while BA changes only slightly. 
This result shows that a sufficient accumulated phase transformation is important for forming a reliable backdoor association.

The three parameters therefore affect different properties of the trigger transformation. 
Based on these results, we use $r=0.9$, $f_c=4000$\,Hz, and $K=10$ as the default configuration throughout the main experiments. 

\subsection{Defense Evaluation}

\subsubsection{Input Preprocessing}

We first evaluate \textsc{GDBack} under four input preprocessing operations: resampling to 8\,kHz, low-pass filtering at 3.4\,kHz, phase jitter, and random all-pass filtering. 
Table~\ref{tab:input_preprocessing} reports the resulting BA and ASR on LibriSpeech for ECAPA-TDNN, x-vector, and RawNet3.

Resampling and low-pass filtering substantially degrade benign recognition, particularly for ECAPA-TDNN and x-vector, while the corresponding ASR remains high. 
For example, low-pass filtering reduces BA to $13.59\%$ and $13.39\%$ on ECAPA-TDNN and x-vector, respectively, whereas ASR remains above $96\%$. 
RawNet3 preserves substantially higher BA under low-pass filtering, but its ASR remains at $98.21\%$. 
These results reveal a substantial tradeoff: aggressive frequency-domain preprocessing markedly reduces benign recognition accuracy while leaving a high target prediction rate on triggered inputs.

Phase jitter and random all-pass filtering exhibit a different behavior. 
Both operations largely preserve BA across the three victim models, but they produce only minor changes in ASR. 
The evaluated preprocessing methods therefore either substantially degrade benign recognition or leave the backdoor largely active, without simultaneously preserving BA and suppressing ASR.

\begin{table*}[t]
\centering
\footnotesize
\caption{Effect of input preprocessing on \textsc{GDBack} on LibriSpeech.}
\label{tab:input_preprocessing}
\renewcommand{\arraystretch}{1.15}
\begin{tabular}{ccccccc}
\toprule
\multirow{2}{*}{Preprocessing} & \multicolumn{2}{c}{ECAPA-TDNN} & \multicolumn{2}{c}{x-vector} & \multicolumn{2}{c}{RawNet3} \\
\cmidrule(lr){2-3} \cmidrule(lr){4-5} \cmidrule(lr){6-7}
& BA (\%) & ASR (\%) & BA (\%) & ASR (\%) & BA (\%) & ASR (\%) \\
\midrule
None                & 98.89 & 99.70 & 99.56 & 98.71 & 99.33 & 94.89 \\
Resample to 8\,kHz   & 10.32 & 91.81 &  9.98 & 78.40 & 55.60 & 96.72 \\
Low pass at 3.4\,kHz & 13.59 & 98.92 & 13.39 & 96.82 & 94.13 & 98.21 \\
Phase jitter        & 98.85 & 99.66 & 99.56 & 98.65 & 99.22 & 94.85 \\
Random APF          & 98.92 & 99.66 & 99.56 & 98.75 & 98.31 & 93.40 \\
\bottomrule
\end{tabular}
\end{table*}

\subsubsection{Neural Cleanse}

We evaluate whether \textsc{GDBack} can be detected by neural cleanse (NC)~\cite{ref22}. 
Neural Cleanse reconstructs class-specific trigger patterns and identifies potential backdoor targets according to the anomaly index of the recovered triggers. 
Following its standard detection criterion, a class is considered suspicious when its anomaly index exceeds 2.

\begin{figure}[t]
\centering
\includegraphics[width=0.8\linewidth]{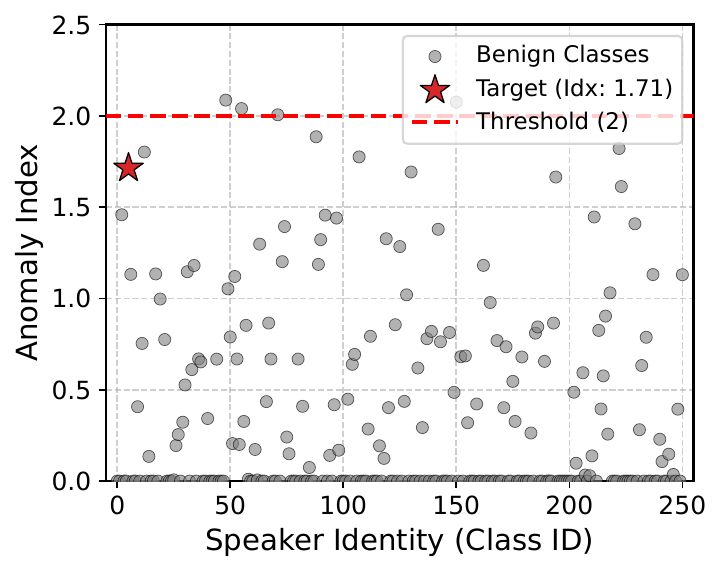}
\caption{Neural Cleanse detection results for \textsc{GDBack}. The actual target class obtains an anomaly index of 1.71, below the detection threshold of 2.}
\label{fig:neural_cleanse}
\end{figure}

As shown in Fig.~\ref{fig:neural_cleanse}, the actual target class obtains an anomaly index of 1.71, which remains below the detection threshold. 
Several benign classes also exhibit comparable or higher anomaly indices, and the true target does not emerge as a distinctive outlier. 
Under this setting, Neural Cleanse therefore fails to identify the backdoor target of \textsc{GDBack}.

\subsubsection{Fine-Tuning}

\begin{figure}[t]
\centering
\subfigure[LibriSpeech]{%
\includegraphics[width=0.48\linewidth]{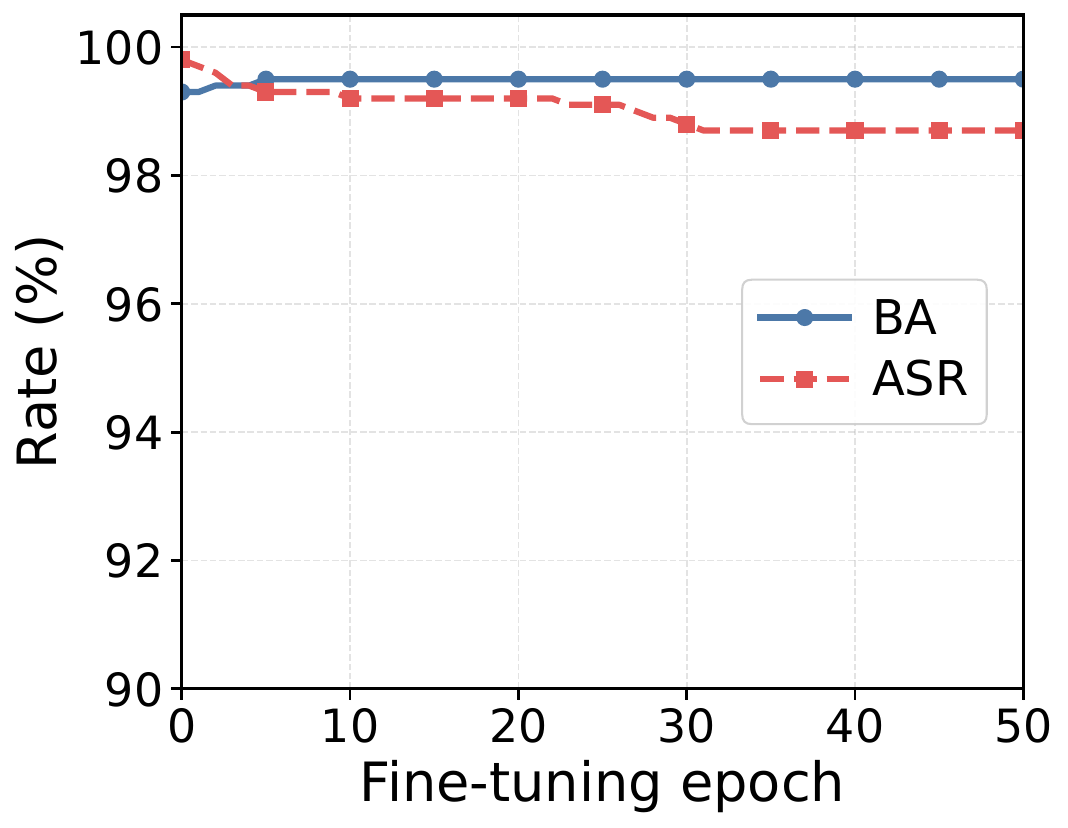}}%
\hfill
\subfigure[VoxCeleb1]{%
\includegraphics[width=0.48\linewidth]{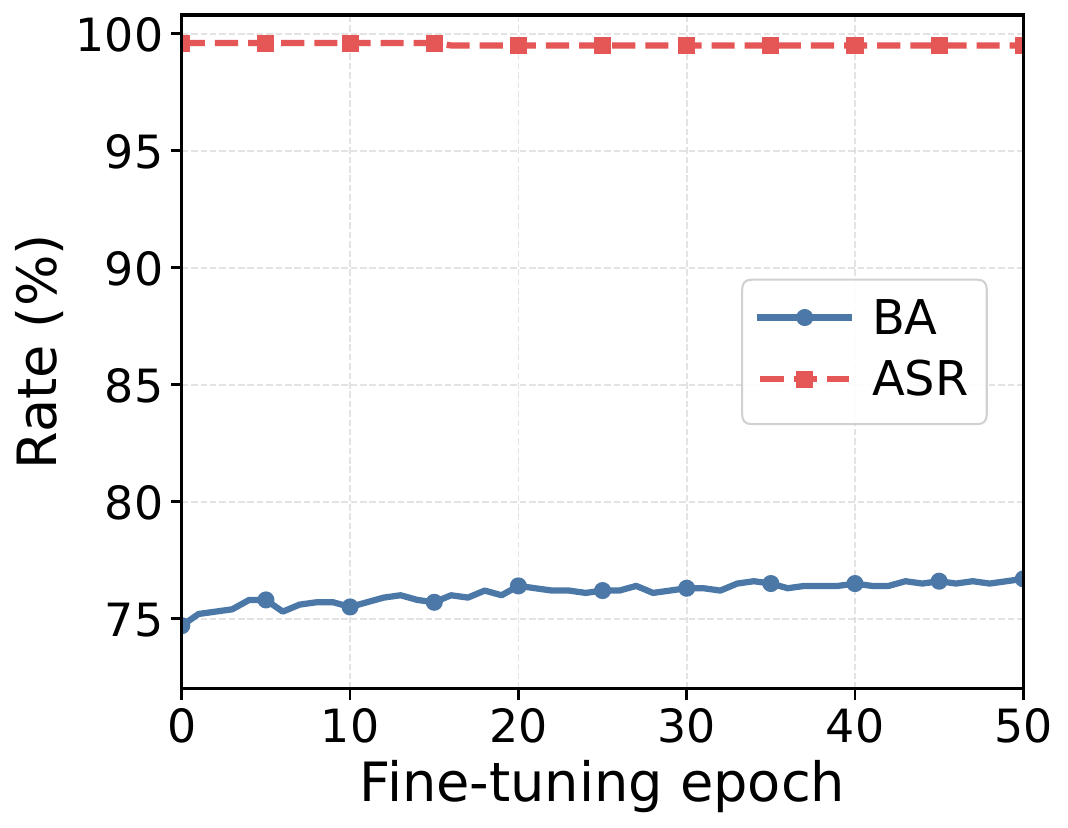}}%

\caption{Effect of clean fine-tuning on \textsc{GDBack} using ECAPA-TDNN. The backdoored model is fine-tuned for 50 epochs using $5\%$ of the clean training data.}
\label{fig:clean_finetuning}
\end{figure}

We evaluate clean fine-tuning as a model repair strategy~\cite{ref42,ref51}. 
For each dataset, we randomly select $5\%$ of the clean training samples and use them to fine-tune the backdoored ECAPA-TDNN for 50 epochs. 
Fig.~\ref{fig:clean_finetuning} reports the evolution of BA and ASR during fine-tuning.

On LibriSpeech, BA remains nearly unchanged throughout the fine-tuning process, while ASR decreases only slightly and stays close to its initial level. 
On VoxCeleb1, BA gradually improves, whereas ASR remains largely unchanged over the 50 epochs. 
These results show that clean fine-tuning under the evaluated setting has only a limited effect on the learned \textsc{GDBack} association.

\subsubsection{Model Pruning}

We further evaluate model pruning as a backdoor mitigation strategy~\cite{ref24}. 
A clean validation set is used to estimate the average activation of neurons in the backdoored ECAPA-TDNN, after which neurons with the lowest average activation are progressively pruned. 
We vary the pruning ratio and report the resulting BA and ASR on LibriSpeech and VoxCeleb1.

\begin{figure}[t]
\centering
\subfigure[LibriSpeech]{%
\includegraphics[width=0.48\columnwidth]{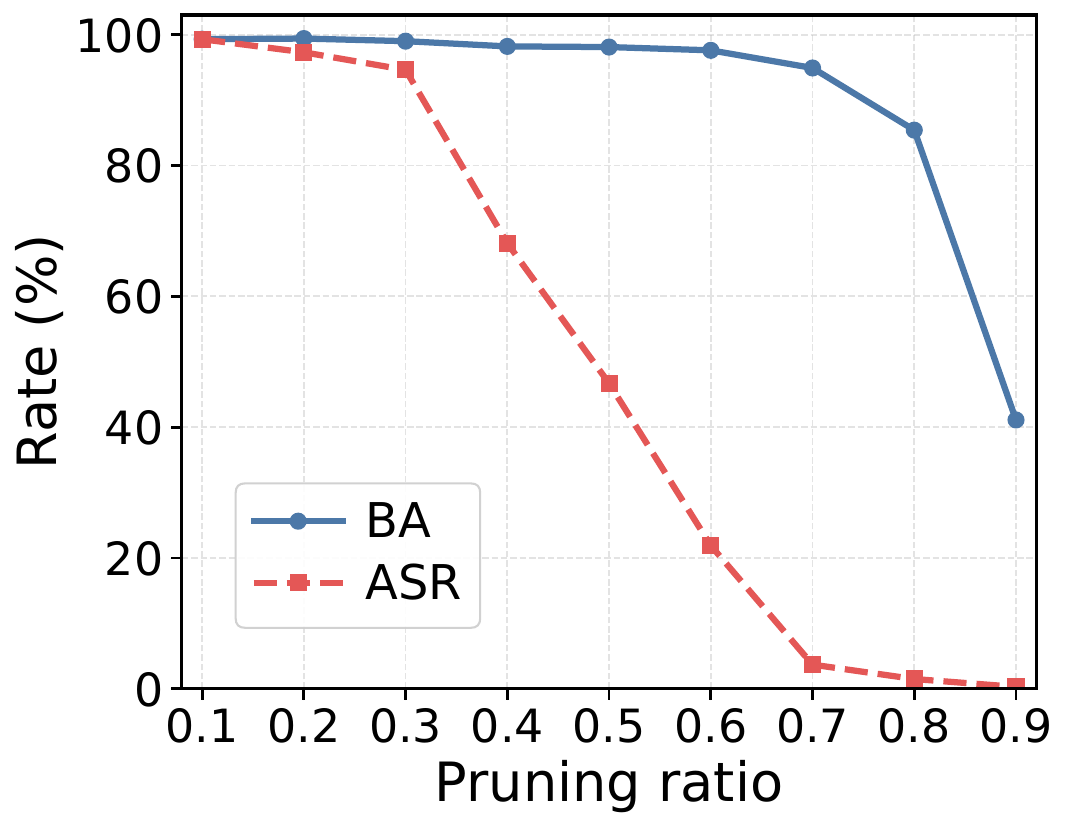}}%
\hfill
\subfigure[VoxCeleb1]{%
\includegraphics[width=0.48\columnwidth]{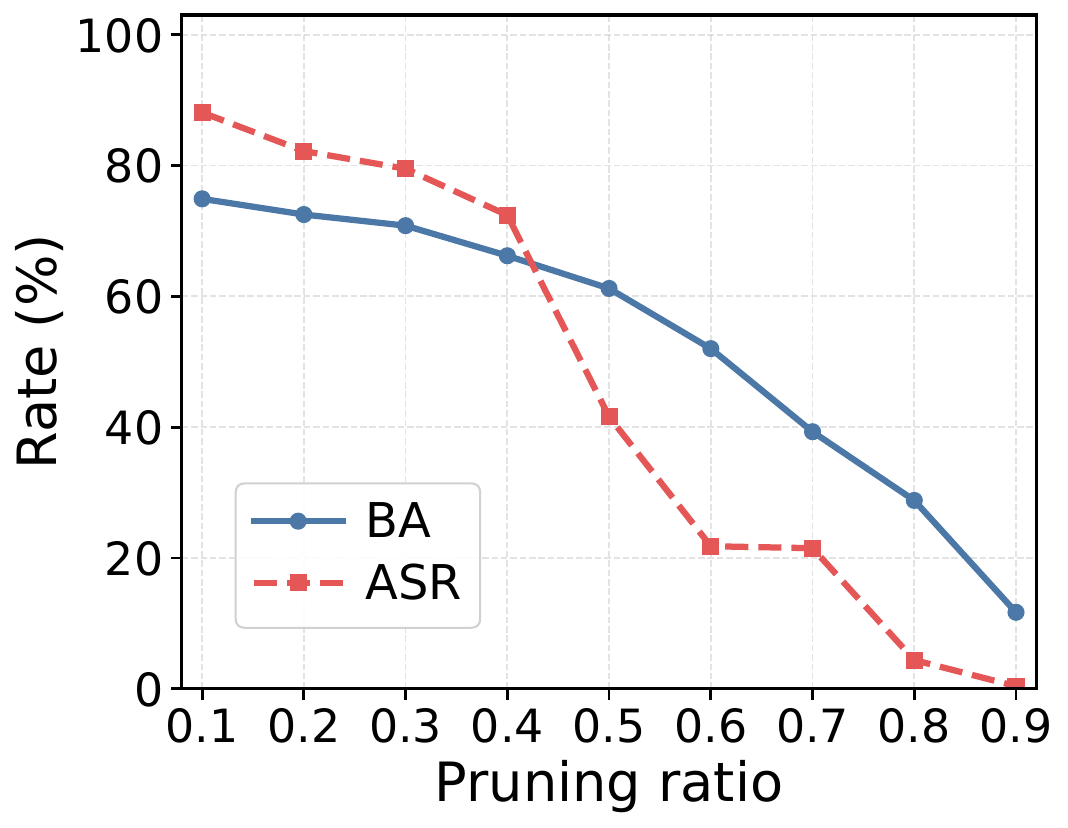}}%

\caption{Effect of model pruning on \textsc{GDBack} using ECAPA-TDNN. 
Neurons with lower average activation on a clean validation set are progressively pruned, and BA and ASR are reported under different pruning ratios.}
\label{fig:model_pruning}
\end{figure}

As shown in Fig.~\ref{fig:model_pruning}, increasing the pruning ratio reduces ASR on both datasets, but the reduction is accompanied by degradation in benign recognition. 
When BA remains close to its original level, the decrease in ASR is limited; more substantial suppression of the backdoor requires stronger pruning, which also causes a noticeable loss in BA. 
These results show that model pruning does not effectively suppress \textsc{GDBack} while preserving benign recognition performance.

\subsection{Over-the-Air Evaluation}

We further evaluate \textsc{GDBack} under real over-the-air (OTA) transmission. 
The experiment is conducted on LibriSpeech using ECAPA-TDNN, x-vector, and RawNet3. 
For each victim model, 200 triggered utterances are played through a smartphone loudspeaker and recorded by a computer placed 1\,m from the playback device. 
The experiment is performed in an ordinary room without acoustic isolation or special soundproofing. 
The recorded utterances are then fed to the corresponding backdoored model. 
We compare the ASR of the digitally triggered speech with that obtained after physical playback and recording.

As shown in Table~\ref{tab:ota_librispeech}, \textsc{GDBack} retains high attack effectiveness after OTA transmission on ECAPA-TDNN and x-vector, with ASRs of $96.50\%$ and $99.00\%$, respectively. 
RawNet3 experiences a larger reduction, with ASR decreasing from $96.00\%$ in the digital setting to $78.00\%$ after playback and recording. 
These results show that, although the degree of degradation varies across victim models, \textsc{GDBack} still maintains a high attack success rate after real acoustic transmission for all three models.

\begin{table}[t]
\centering
\footnotesize
\caption{Over-the-air performance of \textsc{GDBack} on LibriSpeech using 200 triggered utterances for each victim model.}
\label{tab:ota_librispeech}
\renewcommand{\arraystretch}{1.15}
\begin{tabular}{ccc}
\toprule
Model & Digital ASR (\%) & OTA ASR (\%) \\
\midrule
ECAPA-TDNN & 100.00 & 96.50 \\
x-vector   &  98.50 & 99.00 \\
RawNet3    &  96.00 & 78.00 \\
\bottomrule
\end{tabular}
\end{table}

\section{Discussion}

We discuss the implications of transformation-based trigger consistency for phase-based backdoor analysis and defense, as well as the challenges of acoustic transmission and clean-label poisoning.
Backdoor-trigger consistency can reside in the transformation applied to an input rather than in identical features shared across the resulting samples.
This principle underlies speech-transformation attacks and also appears in other modalities, such as the spatial warping mechanism of WaNet \cite{ref54}.
In the phase domain, \textsc{GDBack} introduces the same filter-induced phase response and group-delay contribution across poisoned utterances while preserving their input-dependent output phases.
The sensitivity to filter mismatch, together with the effect of inverse compensation, further suggests that the learned association depends on the specific phase transformation introduced during poisoning.
This perspective offers an alternative to imposing fixed absolute phase templates in phase-based backdoor attacks.

A trigger that is difficult to detect using conventional audio visualizations may still provide a learnable signature.
Mel-spectrograms primarily describe spectral magnitude and thus reveal little about the phase modification introduced by \textsc{GDBack}, while waveform plots also show negligible visual differences between clean and triggered speech.
This discrepancy between visual inspection and model behavior exposes a potential blind spot in magnitude-oriented trigger analysis.
Future defenses could therefore investigate group-delay anomalies and model sensitivity to controlled phase compensation.

Physical playback and recording alter the phase structure of speech, yet the backdoor can still be activated after acoustic transmission.
The greater performance degradation observed for RawNet3 indicates that the evaluated speaker recognition architectures differ in their robustness to acoustic transmission.
Investigating which characteristics of the filter-induced phase response remain effective after transmission could help improve the physical robustness of phase-based backdoors.

In clean-label poisoning, the original speaker labels are preserved \cite{ref20}, so the association must be learned without relabeling the transformed utterances.
Improving the effectiveness of phase-based backdoor attacks under clean-label poisoning remains an important direction for future research.

\section{Conclusion}

We present \textsc{GDBack}, a group-delay backdoor attack for speaker identification that establishes cross-sample trigger consistency through a shared analytic all-pass transformation rather than a fixed absolute phase template.
Across experiments on LibriSpeech and VoxCeleb1 using ECAPA-TDNN, x-vector, and RawNet3, a single trigger configuration achieves consistently high attack success and more consistent performance than the evaluated phase-template baselines while largely preserving benign recognition accuracy.
Controlled interventions indicate that backdoor activation is associated with the induced phase transformation and its filter configuration, while visual comparisons, speech-quality assessments, and speech-content evaluations show that the triggered speech largely retains its original structure, quality, and linguistic content.
The attack also remains effective under the evaluated signal and acoustic variations and after real over-the-air playback and recording, although its performance varies across victim architectures.
Our findings demonstrate that cross-sample trigger consistency in the phase domain can be established through a shared transformation without imposing identical absolute phase patterns, highlighting the importance of phase-aware security analysis for speaker recognition systems.

\balance
\bibliographystyle{IEEEtran}
\bibliography{references}
\end{document}